\documentclass[conference]{IEEEtran}
\IEEEoverridecommandlockouts
\usepackage{cite}
\usepackage{amsmath,amssymb,amsfonts}
\usepackage{epsfig}
\usepackage{algorithmic}
\usepackage{graphicx}
\usepackage{textcomp}
\usepackage{xcolor}
\usepackage{gensymb}
\usepackage{multirow}
\def\BibTeX{{\rm B\kern-.05em{\sc i\kern-.025em b}\kern-.08em
    T\kern-.1667em\lower.7ex\hbox{E}\kern-.125emX}}
\begin{document}

\title{GPS Alignment from Multiple Sources to Extract Aircraft Bearing in Aerial Surveys
\thanks{National Research Council Canada}
}

\author{\IEEEauthorblockN{1\textsuperscript{st} Joshua Power}
\IEEEauthorblockA{\textit{Electrical Engineering} \\
\textit{University of New Brunswick}\\
Fredericton, Canada \\
josh.jgp@unb.ca}
\and
\IEEEauthorblockN{2\textsuperscript{nd} Derek Jacoby}
\IEEEauthorblockA{\textit{Computer Science} \\
\textit{University of Victoria}\\
Victoria, Canada \\
derekja@uvic.ca}
\and
\IEEEauthorblockN{3\textsuperscript{rd} Marc-Antoine Drouin}
\IEEEauthorblockA{\textit{Digital Technologies Research Center} \\
\textit{National Research Council Canada}\\
Ottawa, Canada \\
Marc-Antoine.Drouin@nrc-cnrc.gc.ca}
\and
\IEEEauthorblockN{4\textsuperscript{th} Guillaume Durand}
\IEEEauthorblockA{\textit{Digital Technologies Research Center} \\
\textit{National Research Council Canada}\\
Moncton, Canada \\
Guillaume.Durand@nrc-cnrc.gc.ca}
\and
\IEEEauthorblockN{5\textsuperscript{th} Yvonne Coady}
\IEEEauthorblockA{\textit{Computer Science} \\
\textit{University of Victoria}\\
Victoria, Canada \\
ycoady@uvic.ca}
\and
\IEEEauthorblockN{6\textsuperscript{th} Julian Meng}
\IEEEauthorblockA{\textit{Electrical and Computer Engineering} \\
\textit{University of New Brunswick}\\
Fredericton, Canada \\
jmeng@unb.ca}
}

\maketitle

\begin{abstract}
Methodical aerial population surveys monitoring critically endangered species in Canadian North Atlantic waters are instrumental in influencing government policies both in economic and conservational efforts. The primary factor hindering the success of these missions is poor visibility caused by glare. This paper builds off our foundational paper \cite{cvaui_2022_placeholder} and pushes the envelope toward a data-driven glare modelling system. Said data-driven system makes use of meteorological and astronomical data to assist aircraft in navigating in order to mitigate acquisition errors and optimize the quality of acquired data.  It is found that reliably extracting aircraft orientation is critical to our approach, to that end, we present a GPS alignment methodology which makes use of the fusion of two GPS signals. Using the complementary strengths and weaknesses of these two signals a synthetic interpolation of fused data is used to generate more reliable flight tracks, substantially improving glare modelling. This methodology could be applied to any other applications with similar signal restrictions. 
\end{abstract}

\begin{IEEEkeywords}
GPS alignment, multi-source GPS fusion, aerial survey, marine megafauna, machine learning, glare, data-driven mission planning, context-aware, image quality metric
\end{IEEEkeywords}

% need to add objective, data-driven mission planning using metero, and geospatial data, mitigate errors to optimize quality to of data. Navigation using multiple data sources. Navigate using meteorological and astronomical data.

\section{Introduction}
\label{sec:intro}
Effectively protecting critically endangered marine megafauna like the infamous North Atlantic Right Whale is in part driven by one's ability to provide decision-making entities with accurate population estimates as they are used to inform government policy. North Atlantic Right Whales and other megafauna are systematically surveyed to determine population estimates by efforts coordinated by the Department of Fisheries and Oceans Canada (DFO). Here, Marine Mammal Observers (MMOs) collect data pertaining to identified megafauna and weather conditions. Imagery taken while on survey is used during post-processing to assess meteorological conditions that thwart surface and subsurface visibility, i.e. glare and sea state. 

\begin{figure}[ht]
\begin{center}
\includegraphics[width=.55\linewidth]{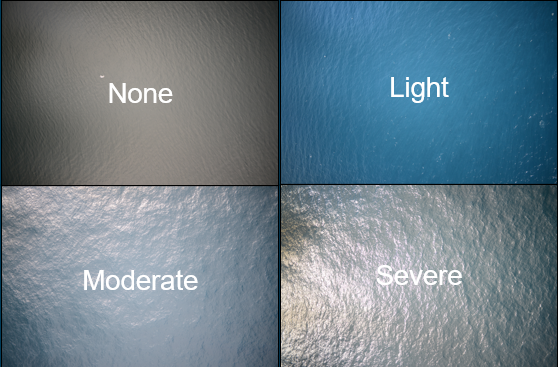} 
\end{center}

   \caption{
   \label{fig:glareEx5}
   Glare intensity classes analyzed by MMOs \cite{powerClassifyingGlareIntensity2021}. \textbf{Top left:} \emph{None} glare class. \textbf{Top right:} \emph{Light} glare class. \textbf{Bottom left:} \emph{Moderate} glare class. \textbf{Bottom right:} \emph{Severe} glare class.
   }
\end{figure}

Post-flight, a detection function is used to fill in gaps where megafaunas are not explicitly seen. This function uses data collected from survey missions to formulate key parameters and is defined in \cite{TNASS2007} by \cite{bucklandIntroductionDistanceSampling2001,bucklandAdvancedDistanceSampling2004,marquesImprovingEstimatesBird2007,marquesIncorporatingCovariatesStandard2003}, where glare is identified as a dominant feature. Glare is separated into four categories which are based on the confidence with which MMOs can detect individuals: \emph{None} – very low chance of missing, \emph{Light} – greater chance of detecting than missing, \emph{Moderate} – greater chance of missing than detecting, and \emph{Severe} – certain to miss some, shown in Figure~\ref{fig:glareEx5}. Classes must be separated in the aforementioned fashion in order to suit the requirements of the detection function. Manually labelling glare intensity in this imagery is labour-intensive and subjective. This leads to a misrepresentation of survey conditions which consequently leads to an incorrectly tuned detection function and an erroneous population estimation.

This project aims to provide population estimates unencumbered by inter-viewer subjectivity. This is to be done by automating the labour-intensive task of post-processing imagery into their respective glare categories. Furthermore, this research intends to develop data-driven mission-planning capacities through the use of multiple geospatial data sources (meteorological and astronomical) in order to plot courses both before and during (real-time updates) flights to assist in mission success. Suggested courses place specific importance on aircraft navigation tuned in order to acquire optimal acquisition data, unencumbered by preventable errors. To that end, we build off our previous line of publications \cite{powerClassifyingGlareIntensity2021, cvaui_2022_placeholder} and begin to answer questions posed in \cite{cvaui_2022_placeholder} with a specific interest in the glare classification error induced by uncertainty in the flight paths. To address this uncertainty we introduce a method to align diverging GPS tracks to extract more reliable aircraft orientation, thereby improving glare classification. %We observe how this change impacts our end results and rejected pseudo-labels. 

% \begin{figure}[ht]
% \begin{center}

%     \includegraphics[width=0.7\linewidth]{gps_img/Flight13_gps.png}
%     \includegraphics[width=0.7\linewidth]{gps_img/Flight13_gps_meta.png}

% \end{center}
%     \caption{\label{fig:aligment_ex} \textbf{Left:} GPS Only. \textbf{Right:} GPS and METADATA.}

% \end{figure}

\section{Literature Review}

% Architectural applications: Not Addressed, not needed.
% Specularity Removal and Detection: Done
% Ocean Remote Sensing: Done
% PV Power Forecasting: Done
% Semi-Supervised Learning: Done
% GPS Alignment: Done

% 3 classes
% CAM3 dataset
% Use of meteorological data

Prior to this manuscript, two glare classification/prediction systems were proposed by this research team. First, a Cascaded Random Forest Model (CRFM) was proposed \cite{powerClassifyingGlareIntensity2021} which uses a small ground-truth dataset (n=294) to classify four classes of glare intensity (\emph{None}, \emph{Light}, \emph{Moderate}, and \emph{Severe}) and achieves a 75.83\% accuracy. The detection function used to infer missed megafauna due to poor visibility required imagery to be classified into the aforementioned four classes, however, this constraint can be ignored when expanding the use case of our system. 

The conditions surrounding our second paper \cite{cvaui_2022_placeholder} permitted the exploration of new research avenues. One is the introduction of meteorological data to provide context-aware insight into physical characteristics affecting the scene understudy. Whilst the other is access to a large unlabeled dataset, i.e. CAM3. Oftentimes, when researchers are given lots of unlabelled data coupled with a small amount of labelled data they'll turn to semi-supervised learning methods to exploit the unlabelled data source as much as possible. Pseudo-labelling \cite{arazoPseudolabelingConfirmationBias2020, shiTransductiveSemisupervisedDeep2018} and consistency regularization \cite{berthelotMixMatchHolisticApproach2019, tarvainenMeanTeachersAre2017} are two dominating semi-supervised learning methodologies used in the machine learning field. Our paper presented a naïve pseudo-labelling methodology to increase available survey data from 294 instances to 98 113 instances. This second iteration of our work focuses on developing a model geared toward glare prediction in order to mitigate poor visibility prior to its occurrence. Since detection function requirements do not need to be consulted for mission planning purposes our glare prediction model restricts classification to three classes where the \emph{None} and \emph{Severe} classes remain and the \emph{Light} and \emph{Moderate} classes are combined into a new \emph{Intermediate} class. Contributions that assisted in informing the selection of features, architectural design, and pre-processing of data for these models and the pipeline presented in this manuscript came from four fields: Glare (specularity or highlight) removal and detection, the ocean remote sensing community, PV Power Forecasting, and GPS Alignment. 

Specularity removal and detection (also referred to as glare removal and detection) is an extensively researched field with relevant insights. Glare removal and detection is typically performed to produce \emph{specular-free} imagery. This can be done by populating specularity maps \cite{artusiSurveySpecularityRemoval2011, khanAnalyticalSurveyHighlight2017} which are proficient in identifying glare intensity quality metrics but they require too many computation resources to support the needs of this research. The concept of the dichromatic reflection model and the associated dichromatic plane is important in understanding how to detect specularity \cite{akashiSeparationReflectionComponents2015,klinkerMeasurementHighlightscolor1988,shaferUsingcolorSeparate1985,shenChromaticitySeparationReflection2008,suoFastHighQuality2016}. Another method, similar to ours, uses a subset of the Hue, Saturation, Intensity (HSI) colour space channel \cite{yangSeparatingSpecularDiffuse2013, zimmerman-morenoAutomaticDetectionSpecular2006}. These methods cannot be directly applied to this research as they do not identify specularity severity, which, for this research, relies on subjective human input. 

The ocean remote sensing community models glare based on the view angle of the observer, solar orientation, and the assumption that capillary waves are representative of ocean waves \cite{Mobley1995,mobleyEstimationRemotesensingReflectance1999}, which, depend on wind speed \cite{coxMeasurementRoughnessSea1954, mobleyLightWaterRadiative1994}. Another study also supports these astronomical importance claims. \cite{mardaljevicDaylightingMetricsThere2012a} found glare to be heavily correlated with the orientation of the sun. It is also found that the sun's orientation can be reliably extracted at any given point near or on Earth's surface \cite{soulaymanCommentsSolarAzimuth2018}.  

Photovoltaic (PV) power forecasting is a closely related field  given the scenes' reliance on solar orientation. A short-term solar forecasting study \cite{wojtkiewiczHourAheadSolarIrradiance2019} using Gated Recurrent Units (GRUs) and Long Short-Term Memory (LSTM) found a significant boost in performance when cloud cover and multivariate deep learning methods were exploited in their methodology. Another investigation observed the effect of using time series neural networks (NNs) to predict solar irradiance using different parameters as input \cite{alzahraniPredictingSolarIrradiance2014}. They found the hour of the day, azimuth and zenith angle of the sun, wind speed, and wind direction to be the most influential features. These findings echo the findings of the ocean sensing community whilst adopting modern techniques. 

The parameters identified as key features by the ocean sensing and PV power forecasting communities are leveraged through meteorological and astronomical data to support data-driven mission planning capabilities in our second model \cite{cvaui_2022_placeholder} which achieves an accuracy of 92.94\% (classification) and 89.22\% (prediction). Similarly, the model presented in this manuscript will also leverage this data. For these context-aware data sources to be adequately utilized the orientation of the aircraft must be reliably extracted, the importance of this information is made clear in Section~\ref{sec:IFM}. To reliably extract aircraft bearing from two diverging GPS signals insight is adopted from geographic information systems (GIS) navigation studies. 

Most GIS navigation methods are aiming to enable autonomous navigation. In order to achieve this, robots or other artificial entities require an accurate map of their environment and current position. Many approaches exist in order to produce reliable navigational services without relying solely on external reference systems like GPS. Simultaneous localization and mapping (SLAM) \cite{dissanayakeComputationallyEfficientSolution2000,gutmannIncrementalMappingLarge1999,hahnelEfficientFastSLAMAlgorithm2003}, GPS/INS (Inertial Navigation Systems) data fusion \cite{gaoMultisensorOptimalData2009,noureldinINSGPSData2004,sasiadekFuzzyAdaptiveKalman2000}, and relative navigation \cite{ellingsonRelativeNavigationFixedwing2020,wheelerRelativeNavigationGPSdegraded2010} are a few popular solutions used to provide reliable localization when GPS signals deviate from their true positions. These approaches use the complementary strengths and weaknesses of different sensors to navigate in diverse conditions. Unfortunately, this project is limited to a single type of sensor, GPS, phasing out SLAM, GPS/INS data fusion, and relative navigation methodologies. Travel time estimation in urban settings is an application which aggregates multiple GPS sources in order to predict travel times and plot more efficient routes. \cite{ehmkeFloatingCarBased2012} and \cite{morgulCommercialVehicleTravel2013} present aggregation approaches that combine homogeneous data in the form of GPS signals along axes of space and time to compute average travel times. Given the challenges posed in urban centers with regard to multi-path errors in GPS signals \cite{kosEffectsMultipathReception2010}, data must be pre-processed to remove nefarious signals. \cite{greavesCollectingCommercialVehicle2008} mitigates these effects by identifying and removing bad data, afterwards these removed points are imputed through logic to recreate points in more appropriate positions. When attempting to determine the locations of individuals through association with GPS-enabled possessions, association confidence is used to best predict the location of individuals \cite{myllymakiLocationAggregationMultiple2002}. This can be explained as weights assigned to certain devices, e.g. phone = 99\%, laptop = 70\%, vehicle = 50\%, etc... in order to guess a person's whereabouts. An analogous method could also be applied to GPS signals with known signal strengths to determine which signals are more reliable and impute points based on the confidence of nearby points.

Narrowing into the specifics of this study, two GPS receivers are relied upon. %one outside of the plane, the other inside, hereby referred to as O-GPS and I-GPS, respectively. 
One type is referred to as I-GPS signals, these are signals from two handheld Garmin GPS devices with antennae situated inside the rear right aircraft bubble. The other type referred to as O-GPS signals, are found outside of a bubble, usually situated in the cockpit. When working with GPS signals it's important to understand errors that might account for uncertainties in GPS data. There are two main types of error: measurement error and sampling error \cite{plaudisAlgorithmicApproachQuantifying2021}. O-GPS requests aircraft position at a fixed rate, making it prone to sampling errors at high speeds. O-GPS signals also tend to be more reliable in regard to measurement error. Whereas, I-GPS positions are acquired at a faster rate than O-GPS and are embedded in image metadata each time an image is acquired. While there are more instances of I-GPS requests, they are often scrambled by the aircraft's cabin making it prone to measurement errors (multi-path effects). Additionally, with this being a marine survey, the surface of the water introduces its own set of challenges causing longer-term multi-path effects when calm waters act like a mirror to incoming GPS signals \cite{kosEffectsMultipathReception2010}. This research has an advantage in that the exact position of the aircraft is of little relevance to us, what's important is reliably extracting aircraft orientation. Therefore, existing flight tracks can be manipulated in order to produce reliable orientation between successive geotags without a string of geotags necessarily being in the correct position. Similar to the GIS navigation approaches discussed, the complementary strengths and weaknesses of O-GPS and I-GPS signals can be used together in order to produce more reliable aircraft headings.

% \cite{jiaUseGPSSensors2004} uses three GPS receivers to appropriately measure aircraft attitude (aircraft orientation relative to the horizon) in order to compensate aircraft effects. 

\section{Image Formation Model}
\label{sec:IFM}
\begin{figure}[ht]
\begin{center}
\includegraphics[width=0.9\linewidth]{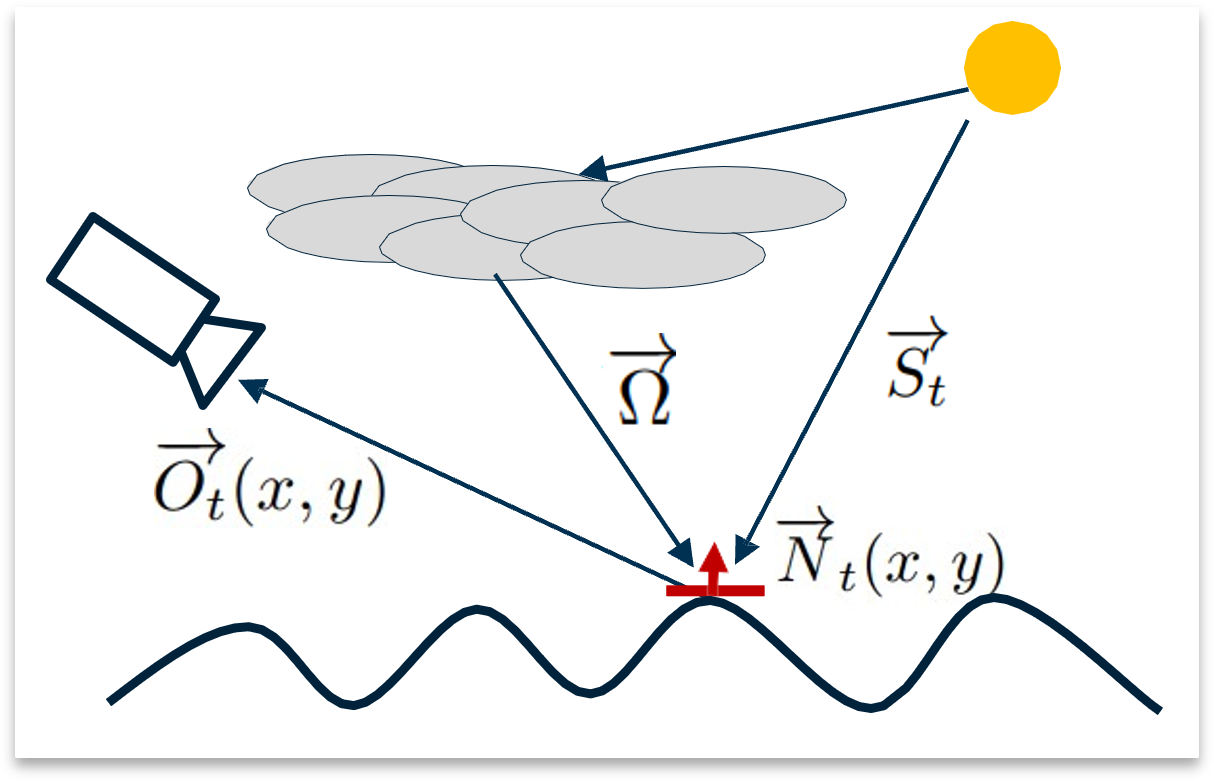}
\end{center}

   \caption{
   \label{fig:glareimgform}
   The image formation model \cite{cvaui_2022_placeholder}, which attempts to encompass the constraints of our real-world system.
%   $ \protect\overrightarrow{O_t}(x,y)$ represent camera orientation with respect a point of the ocean surface, $ \protect\overrightarrow{N}_t(x,y)$ represent the normal of the ocean surface at a given point, $ \protect\overrightarrow{S}_t(x,y)$ represents the orientation of the sun, and $ \protect\overrightarrow{\Omega}$ represents the scenes radiance from the sky i.e. scatter or reflected light from particles in the atmosphere.
   }
\end{figure}

The image formation model introduced in \cite{powerClassifyingGlareIntensity2021,cvaui_2022_placeholder} is illustrated in Figure~\ref{fig:glareimgform}.
The incident angle of pixel $(x,y)$ from a camera at timestamp $t$ with respect to the ground is   $ \overrightarrow{O_t}(x,y)$. 
The sun's relative azimuth and elevation is referred to as  $ \overrightarrow{S_t}$.
The Bidirectional Reflectance Distribution Function (BRDF) of ocean water is denoted $Q$, three vectors are used to define this property, observer viewing angle, light incidence angle, and surface normal. 
 As defined in \cite{powerClassifyingGlareIntensity2021,cvaui_2022_placeholder}, the intensity of the image $I_t$ taken at time $t$ is  
  \begin{multline}
 I_t(x,y) = \\
 R\Big(\int_{2\pi}   Q(  \overrightarrow{O_t}(x,y),\overrightarrow{\Omega}  ,  \overrightarrow{N}_t(x,y)  ) L (\overrightarrow{\Omega}, \overrightarrow{S_t},\overrightarrow{N}_t(x,y)) \, d\overrightarrow{\Omega}\Big)
  \label{eq:brdf}
 \end{multline}
%===============================================
%===============================================
%===============================================
%===============================================
 where $R$ is the response function of the camera, $\overrightarrow{N}_t(x,y)$  is  the normal of the micro-facet of the wave on the ocean surface viewed by pixel $(x,y)$ at time $t$, %
 $\overrightarrow{S_t}$ is the sun's radiance  and cloud coverage\cite{Foley1990}, and $L (\overrightarrow{\Omega}, \overrightarrow{S_t},\overrightarrow{N}_t(x,y))$ represents the scenes radiance from the sky.  
Note that  $\overrightarrow{N}_t(x,y)$ depends on the wind speed $ \overrightarrow{V}$ (see \cite{Mobley1995,mobleyEstimationRemotesensingReflectance1999}).
The severity of the glare at time $t$ is denoted $g_{tc}$ for true classification (4 classes) and $g_{tp}$ for prediction (3 classes). 
$$ g_{tc}  \in \{None,Light,Moderate,Severe \}$$
and
$$ g_{tp}  \in \{None,Intermediate,Severe \}.$$
%Moreover, $g_{tc}$ depends on  $R$, $Q$, $ \overrightarrow{O_t}$ (for all pixels) , $ %\overrightarrow{S_t}$  and $ \overrightarrow{V}$. 
%
%
Explicitly, 
 \begin{equation}
g_{tc} = G_{tc}(I_t,R, \overrightarrow{O_t},Q, \overrightarrow{S_t},  \overrightarrow{V})
\label{eq:gtc}
 \end{equation}
 and
  \begin{equation}
 g_{tp} = G_{tp}(R, \overrightarrow{O_t},Q, \overrightarrow{S_t},  \overrightarrow{V}).
 \label{eq:gtp}
 \end{equation}
 In our previous work, we addressed the determination of $g_{tc}$ and $g_{tp}$ using machine learning tools\cite{powerClassifyingGlareIntensity2021,cvaui_2022_placeholder}. Additionally, $g_{tc}$, is limited to the same three classes as $g_{tp}$ due to pseudo-labelling constraints presented in \cite{cvaui_2022_placeholder}.
Here, Eq.~\ref{eq:brdf} is used to examine the impact of the uncertainties associated with features on model performance.
In order to improve performance, we propose reducing the uncertainties presented by features rather
 then increasing the complexity of the classifier and/or the size of the training data. 
 In the practical implementation of the BRDF, the relative angle between the vector $\overrightarrow{O_t}$, $\overrightarrow{S_t}$, $\overrightarrow{N}_t$, and $\overrightarrow{V}$ are used  \cite{Morel1995,Foley1990}. Thus, any uncertainty introduced by $\overrightarrow{O_t}$ is projected onto other features used by our models. We postulate that a reduction in the uncertainty of $\overrightarrow{O_t}$ would greatly impact model performance. 

\section{Methodology}
\label{sec:methods}
\subsection{Datasets \& Pre-Processing}
\label{sec:data}
% Explain FIT and CAM3 datasets differences. 
% GPS requirements and drops. 
% Provide pseudo-labelling workflow but instruct readers to reference paper. 
% Land-Sea Mask. 

\begin{figure*}[ht]
\begin{center}
\includegraphics[width=.7\linewidth]{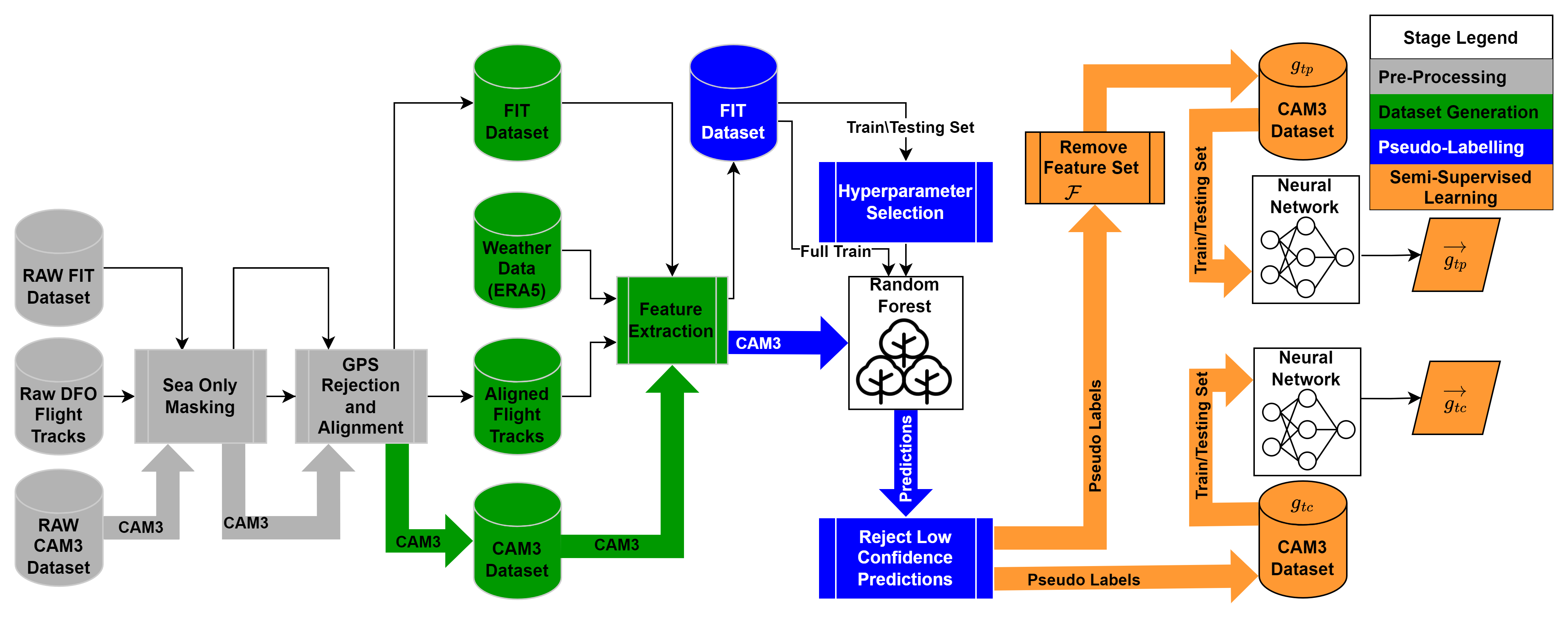} 
\end{center}

   \caption{
   \label{fig:flow} The workflow diagram above is taken from \cite{cvaui_2022_placeholder} but has been altered with the addition of a new pre-processing stage. This diagram illustrates the interactions between all datasets and models in this manuscript. The \textbf{Pre-Processing Stage} highlights how the raw versions of GPS-restricted data sources are first filtered to deduce if instances should be removed from either FIT or CAM3 datasets and combined in order to produce more reliable flight tracks. The \textbf{Dataset Generation Stage} illustrates how datasets are combined to create the resulting features used in the Random Forest and Neural Network architectures employed in this paper. The \textbf{pseudo-labelling stage} depicts how datasets are used to implement the naïve pseudo-labelling approach utilized in this research. The \textbf{Semi-Supervised Learning Stage} exhibits how the pseudo-labelled CAM3 dataset is used for glare classification, $g_{tc}$, and glare prediction, $g_{tp}$.
   }
\end{figure*}

This research has been granted access to two DFO datasets \cite{DFOScienceTwin}: one is nicknamed FIT ($\approx$ 2.7 GB) and the other is nicknamed CAM3 ($\approx$ 2.42 TB). More details can be found in \cite{cvaui_2022_placeholder}. The FIT dataset was labelled by trained MMOs whereas the CAM3 dataset, was pseudo-labelled in accordance with Figure~\ref{fig:flow}, the project pipeline. This naïve pseudo-labelling approach differs slightly from that presented in our previous paper \cite{cvaui_2022_placeholder}, there is now a new pre-processing stage. This has been built in to mitigate orientation errors, more on that in Section~\ref{sec:gps_align}. FIT and CAM3 datasets differ substantially when observing temporal resolution (time between consecutive images). The CAM3 dataset consists of imagery from 33 different flights spanning an average of five hours. The FIT dataset spans 15 different flights and does not consist of enough sequential instances to extract aircraft bearing on its own. The CAM3 dataset acquired imagery at a scattered temporal resolution between 1 and 30 seconds, this is sufficient in determining aircraft bearing in most cases. These flights took place between August and November 2018 between 10:00 and 22:00 (UTC). Geographically, CAM3 consists of data surveyed within latitudes of 41\degree\ to 49\degree\ and longitudes of -55\degree\ to -68\degree.

During pre-processing, four tools are utilized to remove data which might introduce uncertainties when training/testing models. First, a high-resolution sea-land mask is used to filter out all imagery taken overland since imagery containing land masses has been found to negatively impact results \cite{cvaui_2022_placeholder} and is of not interest to this research. Secondly, imagery with no or poor GPS data is removed. Poor GPS signals are quantified as imagery with surrounding geotags (either from I-GPS or O-GPS) greater than 43.75 seconds apart. This number was chosen by observing the turning evolution of the aircraft across all flights, it was found that the average time it took for a 45\degree\ turn was 43.75 seconds. Limiting acceptable imagery to be within this threshold limits sampling error and reduces the chances of undetectable course changes. Thirdly, I-GPS and O-GPS data sources are combined into synthetically interpolated flight tracks as described in Section~\ref{sec:gps_align} in order to generate a more reliable orientation metric. Lastly, the remaining data is distributed into the pseudo-labelling procedure where low-confidence predictions are rejected prior to training/testing. In \cite{cvaui_2022_placeholder} 98 113 pseudo-labels remain after pre-processing. Using the tools described in this manuscript results in 76 102 pseudo-labels. 77 851 were accepted after pseudo-labelling but 1749 were taken over-land. Of these 76 102 pseudo-labels, 53 125 were from the \textit{None} class, 14 900 were from the \textit{Intermediate} class, and 8077 were from the \textit{Severe} class.

% should mention how much data remained after pre-processing for both datasets. 15 tracks remain for CAM3.

\subsection{GPS Alignment}
\label{sec:gps_align}
\begin{figure}[ht]
\begin{center}
\fbox{\includegraphics[width=.45\linewidth]{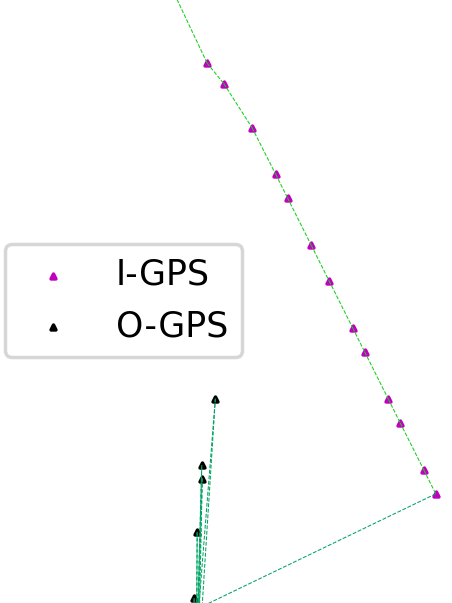}}
\fbox{\includegraphics[width=.45\linewidth]{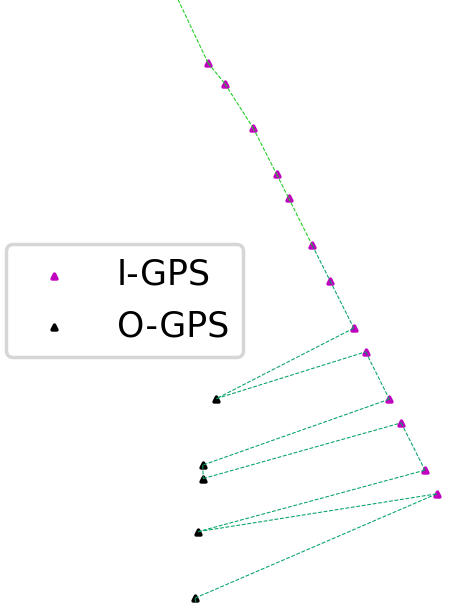}}\\
\vspace{0.1cm}
\fbox{\includegraphics[width=.95\linewidth]{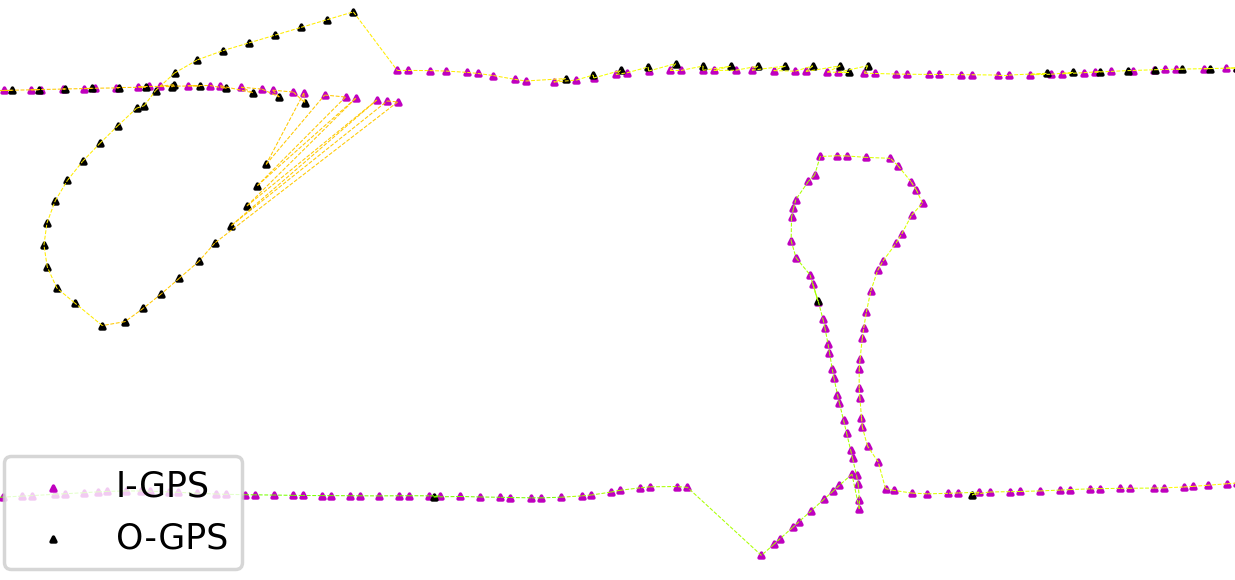}}
\end{center}

   \caption{
   \label{fig:offset} \textbf{Top Left:} Flight tracks before offset is applied. \textbf{Top Right:} Flight tracks after offset is applied. \textbf{Bottom:} Post offset larger section.
   }
\end{figure}

The GPS alignment methodology described in this manuscript is designed to utilize the complementary strengths of O-GPS and I-GPS signals in order to produce flight tracks that are more reliable at extracting aircraft orientation. While observing the differences between these two signals it was found that O-GPS and I-GPS signals had significant temporal offsets which made data fusion difficult when extracting bearing. To combat this nearest O-GPS and I-GPS signals were deduced by parsing through geotags and gathering points within $\pm$60 seconds and computing euclidean distances to points of the opposite species. The smallest euclidean distances and their respective offsets are shown in Table~\ref{tab:offsets}. Then respective offsets were applied to individual flights as the first step in our alignment algorithm; an illustration of this step is shown in Figure~\ref{fig:offset}.

\begin{table}[ht]
\caption{\label{tab:offsets} I-GPS, O-GPS Temporal Offsets on a Per-Flight Basis}
\begin{center}     
\begin{tabular}{|l|c|c|} %% this creates two columns
%% |l|l| to left justify each column entry
%% |c|c| to center each column entry
%% use of \rule[]{}{} below opens up each row
\hline
\multicolumn{3}{|c|}{ \textbf{Per-Flight Offsets}} \\ 
\hline
\cline{1-3}
	\textbf{Flight Num.} &\textbf{Euclidean Dist. (m)} &\textbf{Offset (Seconds)  }\\
  	 \cline{1-3}
1	&0.746	  &-5 \\
\cline{1-3}
2	&4.113    &-5 \\
\cline{1-3}
3	&4.901	  &-6 \\
\cline{1-3}
4	&58.35	  &-5 \\
\cline{1-3}
5	&2.636	  &-5 \\
\cline{1-3}
6	&2.200	  &-4 \\
\cline{1-3}
7	&5.551	  &-5 \\
\cline{1-3}
8	&1.129	  &-5 \\
\cline{1-3}
10	&3.179	  &-5 \\
\cline{1-3}
11	&4.000	  &-4 \\
\cline{1-3}
12	&4.352	  &-8 \\
\cline{1-3}
13	&3.009	  &-3 \\
\cline{1-3}
17	&0.934	  &0  \\
\cline{1-3}
18	&36.87	  &0  \\
\cline{1-3}
19	&0.735	  &2  \\
\cline{1-3}
21	&0.758	  &-3 \\
\cline{1-3}
22	&4.040	  &1  \\
\cline{1-3}
23	&0.771	  &2  \\
\cline{1-3}
24	&0.753	  &4  \\
\cline{1-3}
25	&0.394	  &7  \\
\cline{1-3}
29	&3.205	  &17 \\
\cline{1-3}
30	&0.395	  &16 \\
\cline{1-3}
31	&2.896	  &15 \\
\cline{1-3}
32	&0.657	  &16 \\ 
\hline
\end{tabular}

\end{center}

\end{table}  

\begin{figure}[ht]
\begin{center}
\fbox{\includegraphics[width=.95\linewidth]{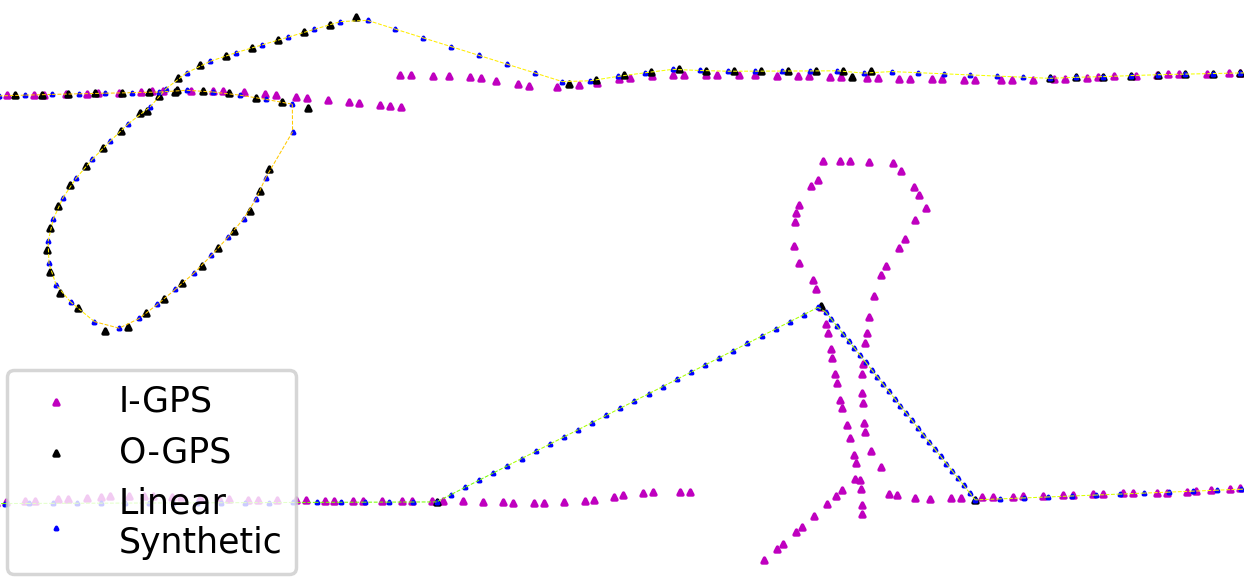}}
\end{center}

   \caption{
   \label{fig:linear} Linear synthetic interpolation at 5-second temporal resolution.
   }
\end{figure}

Next, all geotags (O-GPS and I-GPS) from a given flight are converted from latitude/longitude coordinate systems into a 3D Cartesian plane. A linear synthetic interpolation is performed along the x and y axes of O-GPS signals to create a linear synthetic interpolation, each synthetic point is spaced 5 seconds apart, shown in Figure~\ref{fig:linear}. For each linear synthetic point, $S_i$, $M$ surrounding points are gathered if they fall within $\pm$40 seconds of the synthetic timestamp. For each of these surrounding points, $P_n$, with coordinates $(X_n, Y_n)$, that falls within this threshold a corresponding weight term, $W_n$, is computed and used in accordance with Equations~\ref{eq:moving_average_x} and~\ref{eq:moving_average_y}. These equations describe how points are interpolated along the x and y axes where a weighted synthetically interpolated point, $SI_i$, has coordinates $(XSI_i, YSI_i)$.

 \begin{equation}
XSI_i = \frac{\sum_{n=0}^{M} {W_n*X_n}}{\sum_{n=0}^{M} {W_n}}
\label{eq:moving_average_x}
 \end{equation}
 
  \begin{equation}
YSI_i = \frac{\sum_{n=0}^{M} {W_n*Y_n}}{\sum_{n=0}^{M} {W_n}}
\label{eq:moving_average_y}
 \end{equation}

\begin{figure}[ht]
\begin{center}
\fbox{\includegraphics[width=.95\linewidth]{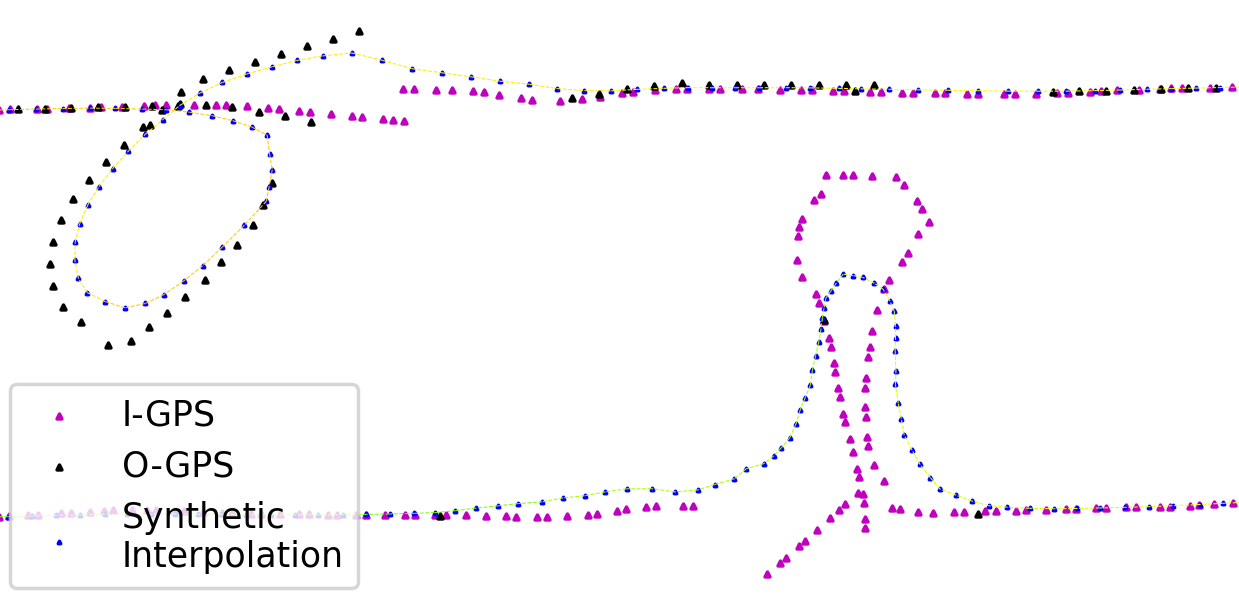}}

\end{center}

   \caption{
   \label{fig:interpolated} Moving average synthetic interpolation used for reliable aircraft bearing extraction. \textbf{Note:} This figure is a small section of a single flight.
   }
\end{figure}

Similar to association confidence weightings \cite{myllymakiLocationAggregationMultiple2002} our alignment formula weights signals differently. O-GPS signals are known to be more reliable than I-GPS signals, therefore their weight should be higher when trying to deduce new interpolated points. Equation~\ref{eq:weights} shows how a given point, $P_n$, is assigned its weighting term, $W_n$. The source weight, $W_{source}$, is determined based on the signal type of $P_n$, there are three options. O-GPS, I-GPS, or Synthetic, each is assigned values of 8, 5, and 5 respectively. The temporal weight, $W_{temporal}$, is assigned a value of 0.5. The source weight terms and temporal weight term were tuned through trial and error until results looked favourable across all flights. Lastly, the time difference, $T_{diff}$, between the point of interest, $S_i$, and a given point $P_n$ is determined in units of seconds. The resulting synthetic moving average interpolation is shown in Figure~\ref{fig:interpolated}. 

 \begin{equation}
W_n = \frac{W_{source}}{W_{temporal}*|T_{diff}|+1}
\label{eq:weights}
 \end{equation}

\subsection{Feature Selection \& Extraction}
\label{sec:features}
Deep-learning methods such as convolutional neural networks (CNNs) are known to extract application-specific features during training \cite{joginFeatureExtractionUsing2018}. However, given the data-hungry requirements of deep learning methods like CNNs, and the modest size of the FIT dataset, we extract our own small set of image-related features. The feature set ${\cal F}$ consists of six imagery-specific features extracted from the histogram of the I channel of the HSI colour space of a given image $I_t$. To remain consistent across datasets the feature extraction process for ${\cal F}$ is identical in FIT and CAM3 datasets even though CAM3 is much larger. ${\cal F}$ comprises of the features: \textit{Saturation}, \textit{Max} \textit{Index}, \textit{Max}, \textit{Skew}, and \textit{Width}. All features except for \textit{Width} are described in \cite{powerClassifyingGlareIntensity2021}, whereas, \textit{Width} is described in \cite{cvaui_2022_placeholder}. 

The features that remain are all extracted from metadata, i.e. latitude, longitude, altitude, and timestamp. With the use of this metadata the position of the sun with respect to the aircraft, $\overrightarrow{S_t}$, can be determined and is referred to using two coordinates as \textit{Azimuth} and \textit{Alt Degree}. Knowing the orientation of the aircraft, \textit{bearing}, from methods introduced in Section~\ref{sec:gps_align}, one can normalize the positions of each camera with respect to $\overrightarrow{S_t}$. This is done to determine $\overrightarrow{O_t}$, the orientation of the camera with respect to a point on the surface of the ocean, referred to as \textit{Azimuth absolute diff}. All meteorological features are represented by the feature set $\overrightarrow{M}$, these are features which pertain to wind ($\overrightarrow{V}$), sea state (approximation of $\overrightarrow{N_t}$), and cloud cover. The meteorological data used to inform these features were downloaded from the Copernicus Climate Data Store \cite{hersbachERA5HourlyData2018}. This source uses ERA5 (the 5th generational European Center for Medium-Range Weather Forecasts) to reanalyze historical meteorological forecasts. The feature set $\overrightarrow{M}$ consists of the following weather variables: \textit{Maximum individual wave height}, \textit{Mean direction of total swell}, \textit{Mean direction of wind waves},\textit{ Mean period of total swell}, \textit{Mean period of wind waves}, \textit{Mean wave direction}, \textit{Mean wave period}, \textit{Significant height of combined wind waves and swell}, \textit{Significant height of wind waves}, \textit{High cloud cover}, \textit{Low cloud cover}, \textit{Medium cloud cover}, \textit{Total cloud cover}, \textit{Cloud above aircraft}. \textit{Cloud above aircraft} is the only variable not available from Copernicus as it is computed on a per-image basis as a binary feature indicating whether or not there is a cloud formation above the coordinate of interest. Additionally, all directional data variables have sister terms which are normalized with respect to the camera and have \textit{absolute diff} added to their feature name, eg. \textit{Mean direction of wind waves} becomes \textit{Mean direction of wind waves absolute diff}. In this manuscript, normalized sister terms are used, not their unnormalized counterparts. Altogether these are the 27 features used as inputs to our classification model, $g_{tc}$. However, for our prediction model, $g_{tp}$, we remove the feature set $\cal F$ and are left with 22 features unaccompanied by imagery-related features. 

\section{Experimental Results}
\label{sec:experimental}
\subsection{MLP Results}
\label{sec:mlp classify}
The same NN architectures and hyper-parameters as our previous work \cite{cvaui_2022_placeholder} are used in this manuscript to compare performance enhancements between GPS-Aligned and Non-GPS-Aligned datasets. Said network architectures are MLPs, densely connected between all layers, one is used for classification, $g_{tc}$, and the other is used for prediction, $g_{tp}$. The input layer consists of 27 nodes ($\mathbb{R}^{27}$, $g_{tc}$) and 22 nodes ($\mathbb{R}^{22}$, $g_{tp}$), both hidden layers consist of 256 nodes ($\mathbb{R}^{256}$), and the output layer is made up of 3 nodes ($\mathbb{R}^{3}$). The pseudo-labelled CAM3 dataset undergoes 5-fold cross-validation, with each training/testing fold containing similar class proportions. Additionally, each training and testing fold undergoes normalization separately as defined by Equation~\ref{eq:normalize}. Cross-entropy loss is used to adjust model weights with loss function weights adjusted in accordance with Equation~\ref{eq:balance}. This is done to combat class imbalance by forcing weights in the loss function to be inversely proportional to class frequency in the training set.

\begin{equation}
\overrightarrow{F_{norm}}=\frac{\overrightarrow{F_{col}} - min(\overrightarrow{F_{col}})}{max(\overrightarrow{F_{col}}) - min(\overrightarrow{F_{col}}) + 1e^{-6}}
\label{eq:normalize}
\end{equation}

\begin{equation}
W_i=\frac{\#_{samples}}{\#_{classes}\#_{instances}}
\label{eq:balance}
\end{equation}

An Edisonian process guided by monitoring training/testing loss and accuracy informed the selection of hyper-parameters. To ascertain these parameters the following was explored: hidden layer depths of 1 and 2, hidden layer nodes between 8 and 256, epochs between 40 and 300, learning rates between $1e^{-7}$ and 0.1, both LBFGS (Limited-memory Broyden-Fletcher-Goldfard-Shanno algorithm) and Adam optimizers were explored. MLP classification performance, $g_{tc}$ is shown in Table~\ref{tab:res_class}, whereas prediction performance, $g_{tp}$, is shown in Table~\ref{tab:res_pred}. Hyper-parameters for each of these models are shown in the lower portion of these tables.

\begin{table}[ht]
\caption{\label{tab:res_class}MLP Classification Performance, $g_{tc}$}
\begin{center}       
\begin{tabular}{|l|c|c|c|c|c|} %% this creates two columns
%% |l|l| to left justify each column entry
%% |c|c| to center each column entry
%% use of \rule[]{}{} below opens up each row
\hline
\multicolumn{5}{|c|}{ \textbf{Imagery \& Metadata MLP (No GPS Alignment)}} \\ 
\hline
\cline{1-5}
	\textbf{(n = 98113)} &Precision	&Recall	&F-Score   &Accuracy  \\
  	 \cline{1-5}
None	&95.82	&91.72	&93.73	&  \multirow{3}{*}{92.94}	\\
\cline{1-4}
Intermediate		&86.81	&92.72	&89.67    &	        \\
\cline{1-4}
Severe		&98.07	&99.47	&98.77    &	      \\
\hline
\multicolumn{5}{|c|}{ \textbf{Imagery \& Metadata MLP (Proposed GPS Alignment)}} \\ 
\hline
\cline{1-5}
	\textbf{(n = 76102)} &Precision	&Recall	&F-Score   &Accuracy  \\
  	 \cline{1-5}
\textbf{None}	           &99.02      &95.57      &97.27      &\multirow{3}{*}{96.22}	\\
\cline{1-4}
\textbf{Intermediate}	   &86.03      &96.47      &90.94      &	        \\
\cline{1-4}
\textbf{Severe}		       &99.56      &99.91      &99.74      &	        \\
\hline

\multicolumn{5}{|c|}{ \textbf{Classification Hyper-Parameters}} \\ 
\hline
\cline{1-5}
\multicolumn{2}{|l|}{\textbf{Learning}}        &		    &\textbf{Hidden}	        &\\
\multicolumn{2}{|l|}{\textbf{Rate}}            &0.01          &\textbf{Layer}             & ReLU \\
 \multicolumn{2}{|l|}{}                        &               &\textbf{Activation}        & \\
\hline
\multicolumn{2}{|l|}{\textbf{Hidden}}	    &[256, 256]	   &\textbf{Batch Size}         &[8192, 8192]	        \\
\multicolumn{2}{|l|}{\textbf{Layers}}       &              &                            &                       \\
\hline
\multicolumn{2}{|l|}{\textbf{Epochs}}		        &50            &\multirow{2}{*}{\textbf{Norm. Data}} &\multirow{2}{*}{Yes}	      \\
\cline{1-3}
\multicolumn{2}{|l|}{\textbf{Optimizer}}            &Adam         &                                     &\\
\hline

\end{tabular}
\end{center}

\end{table}  

\begin{table}[ht]
\caption{\label{tab:res_pred}MLP Prediction Performance, $g_{tp}$}
\begin{center}       
\begin{tabular}{|l|c|c|c|c|} %% this creates two columns
%% |l|l| to left justify each column entry
%% |c|c| to center each column entry
%% use of \rule[]{}{} below opens up each row
\hline
\multicolumn{5}{|c|}{ \textbf{Metadata Only MLP (No GPS Alignment)}} \\ 
\hline
\cline{1-5}
	\textbf{(n = 98113)} &Precision	&Recall	&F-Score   &Accuracy  \\
  	 \cline{1-5}
None	&94.89	&88.75	&91.72	&  \multirow{3}{*}{89.22}	\\
\cline{1-4}
Intermediate		&82.82	&88.59	&85.61    &	        \\
\cline{1-4}
Severe		&83.61	&93.29	&88.18    &	      \\
\hline
\multicolumn{5}{|c|}{ \textbf{Metadata Only MLP (Proposed GPS Alignment)}} \\ 
\hline
\cline{1-5}
	\textbf{(n = 76102)} &Precision	&Recall	&F-Score   &Accuracy  \\
  	 \cline{1-4}
\textbf{None}	           &98.90      &93.49      &96.12      &\multirow{3}{*}{94.01}	\\
\cline{1-4}
\textbf{Intermediate}	   &81.56      &94.84      &87.70      &	        \\ 
\cline{1-4}
\textbf{Severe}		       &90.54      &95.87      &93.13      &	        \\
\hline

\multicolumn{5}{|c|}{ \textbf{Prediction Hyper-Parameters}} \\ 
\hline
\cline{1-5}
\multicolumn{2}{|l|}{\textbf{Learning}}        &		    &\textbf{Hidden}	        &\\
\multicolumn{2}{|l|}{\textbf{Rate}}            &0.01          &\textbf{Layer}             & ReLU \\
 \multicolumn{2}{|l|}{}                        &               &\textbf{Activation}        & \\
\hline
\multicolumn{2}{|l|}{\textbf{Hidden}}	    &[256, 256]	   &\textbf{Batch Size}         &[8192, 8192]	        \\
\multicolumn{2}{|l|}{\textbf{Layers}}       &              &                            &                       \\
\hline
\multicolumn{2}{|l|}{\textbf{Epochs}}		        &50            &\multirow{2}{*}{\textbf{Norm. Data}} &\multirow{2}{*}{Yes}	      \\
\cline{1-3}
\multicolumn{2}{|l|}{\textbf{Optimizer}}            &Adam         &                                     &\\
\hline
\end{tabular}

\end{center}

\end{table}  

\subsection{Rejection and SHapley Analyses}
\label{sec:shap_reject}

\begin{figure}[ht]
\begin{center}
\fbox{\includegraphics[width=.95\linewidth]{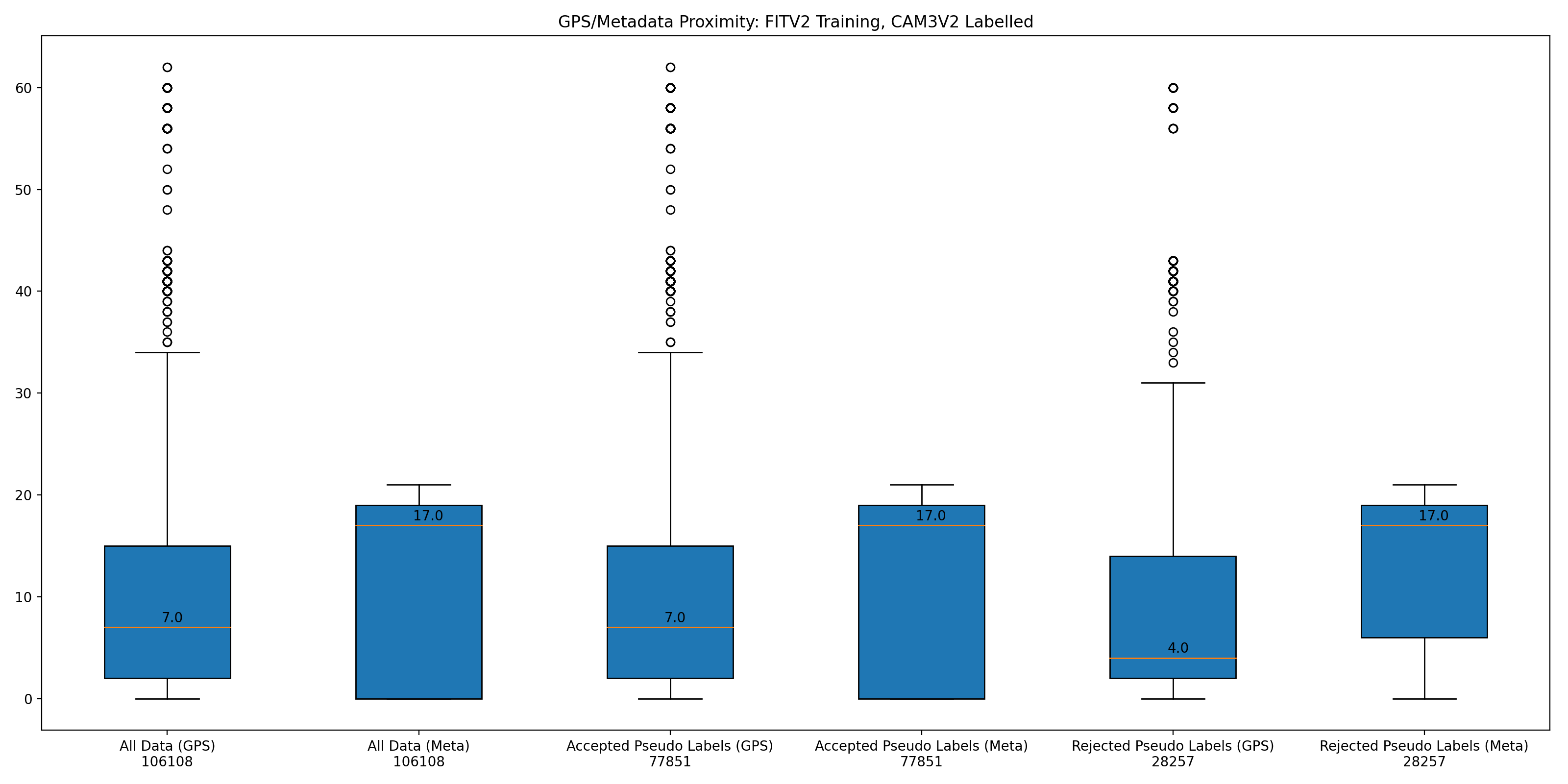}}
\end{center}

   \caption{
   \label{fig:reject} Rejection analysis box plot. GPS refers to O-GPS signals and Meta refers to I-GPS signals.
   }
\end{figure}

As outlined in our previous paper \cite{cvaui_2022_placeholder} a preliminary investigation into parameters that might impact pseudo-label rejection yielded insufficient evidence for concrete conclusions. This was believed to be due to unreliable aircraft orientation metrics, hence, the use of the GPS alignment methodology presented in this manuscript. A more comprehensive rejection analysis took place after GPS alignment was applied. Specifically, an analysis of data used to inform features that are less reliably extracted. Given the low resolution of meteorological data (0.25\degree\ by 0.25\degree\ at 1-hour increments) a specific investigation into meteorological effects on rejection was pursued. To achieve this three groups of data were identified, one consisted of all data, the second consisted of all accepted pseudo-labels, and the third consisted of all rejected labels. Each of these data points was made up of a timestamp, latitude, and longitude and each was assessed on its spatiotemporal proximity to the nearest available meteorological points. Unfortunately, this investigation saw no difference between clustered groups. However, Figure~\ref{fig:reject} represents what occurs when the same analysis was employed to observe the number of O-GPS and I-GPS points within $\pm$30 seconds of points of interest. 30 seconds was chosen as it is between 43.75 seconds (the average 45\degree\ turn time) and 20.75 seconds (the fastest recorded 45\degree\ turn time). 

\begin{figure*}[ht]
\begin{center}
\fbox{\includegraphics[width=.95\linewidth]{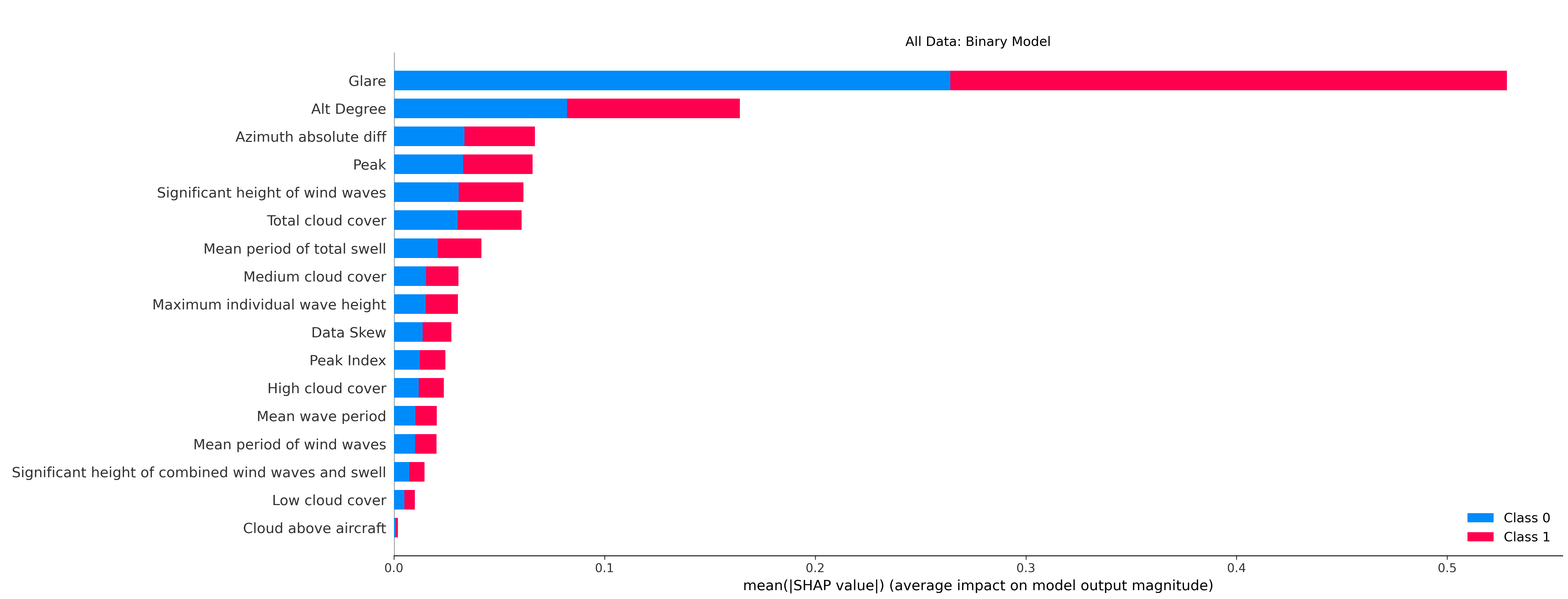}}\\
\vspace{0.1cm}
\fbox{\includegraphics[width=.95\linewidth]{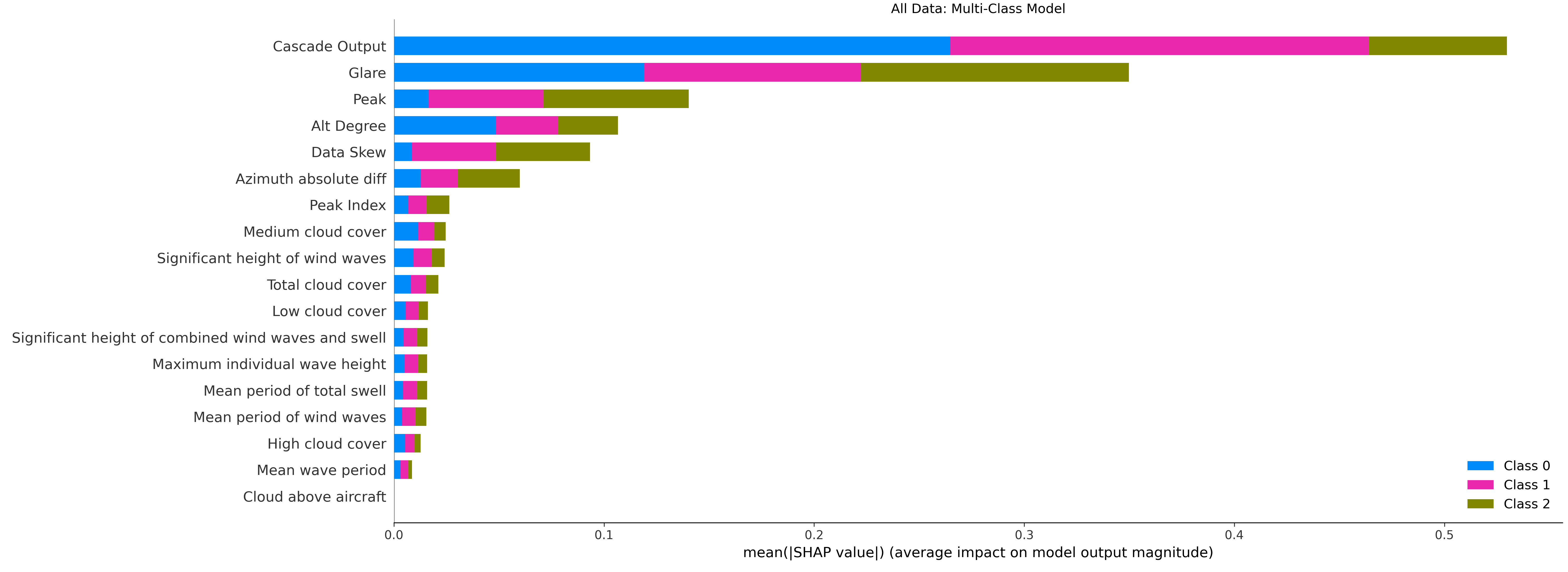}}
\end{center}

   \caption{
   \label{fig:CRFM_shap} Summary plot of SHapely values for CRFM \textbf{(Proposed GPS Alignment)}. \textbf{Top}: Binary Classification Stage, where Class: 0 is \textit{None} and Class: 1 is \textit{Intermediate} and \textit{Severe}. \textbf{Bottom:} Final Classification Stage, where Class: 0 is \textit{None}, Class: 1 is \textit{Intermediate}, and Class: 2 is \textit{Severe}. 
   }
\end{figure*}

\begin{figure*}[ht]
\begin{center}
\fbox{\includegraphics[width=.95\linewidth]{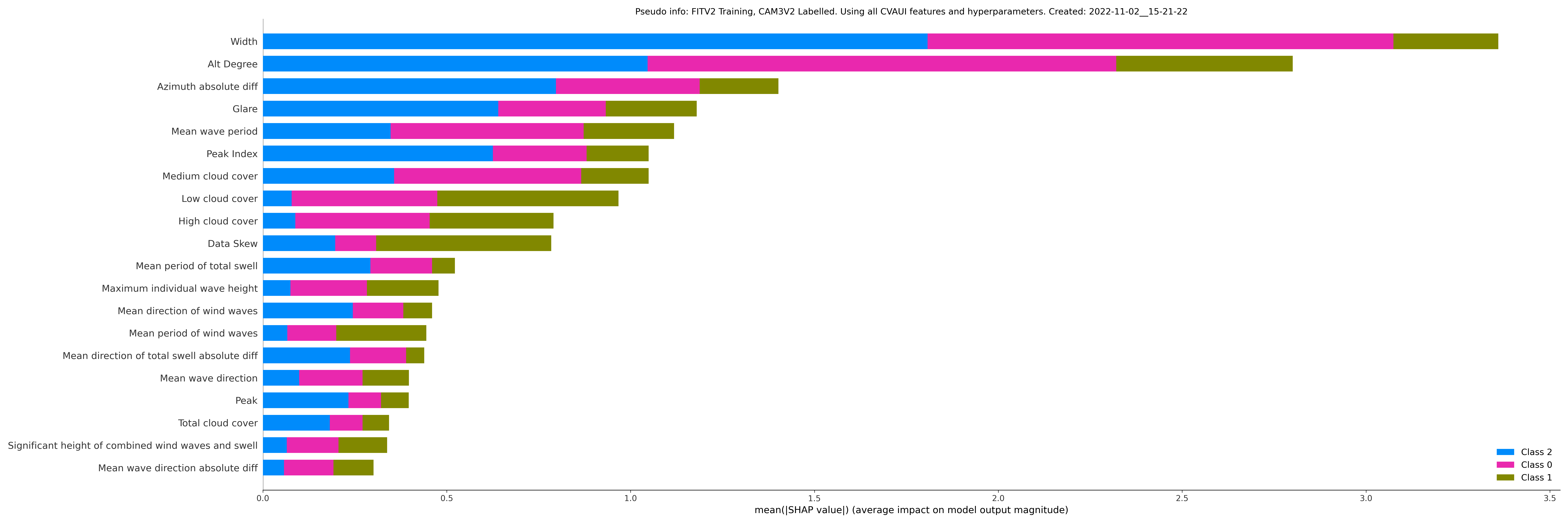}}\\
\vspace{0.1cm}
\fbox{\includegraphics[width=.95\linewidth]{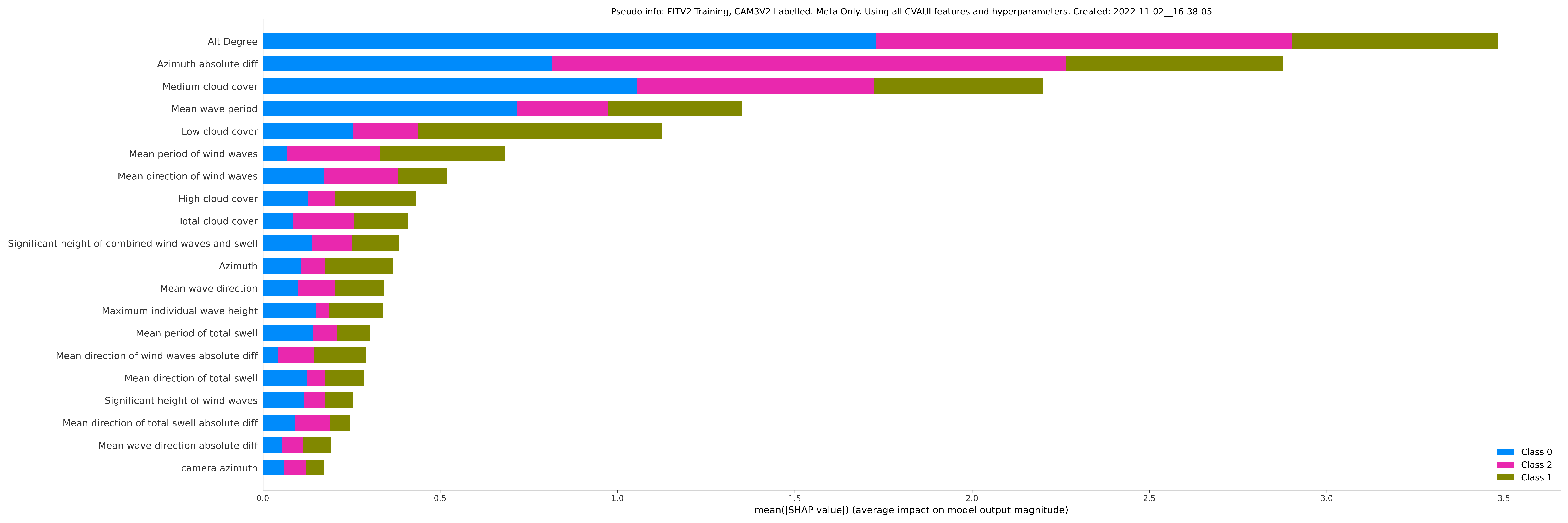}}
\end{center}

   \caption{
   \label{fig:MLP_shap} Summary plot of SHapely values for MLP \textbf{(Proposed GPS Alignment)}. \textbf{Top:} Classification MLP, $g_{tc}$, where Class: 0 is \textit{None}, Class: 1 is \textit{Intermediate}, and Class: 2 is \textit{Severe}. \textbf{Bottom:} Prediction MLP, $g_{tp}$, where Class: 0 is \textit{None}, Class: 1 is \textit{Intermediate}, and Class: 2 is \textit{Severe}.
   }
\end{figure*}

\begin{figure*}[ht]
\begin{center}
\fbox{\includegraphics[width=.95\linewidth]{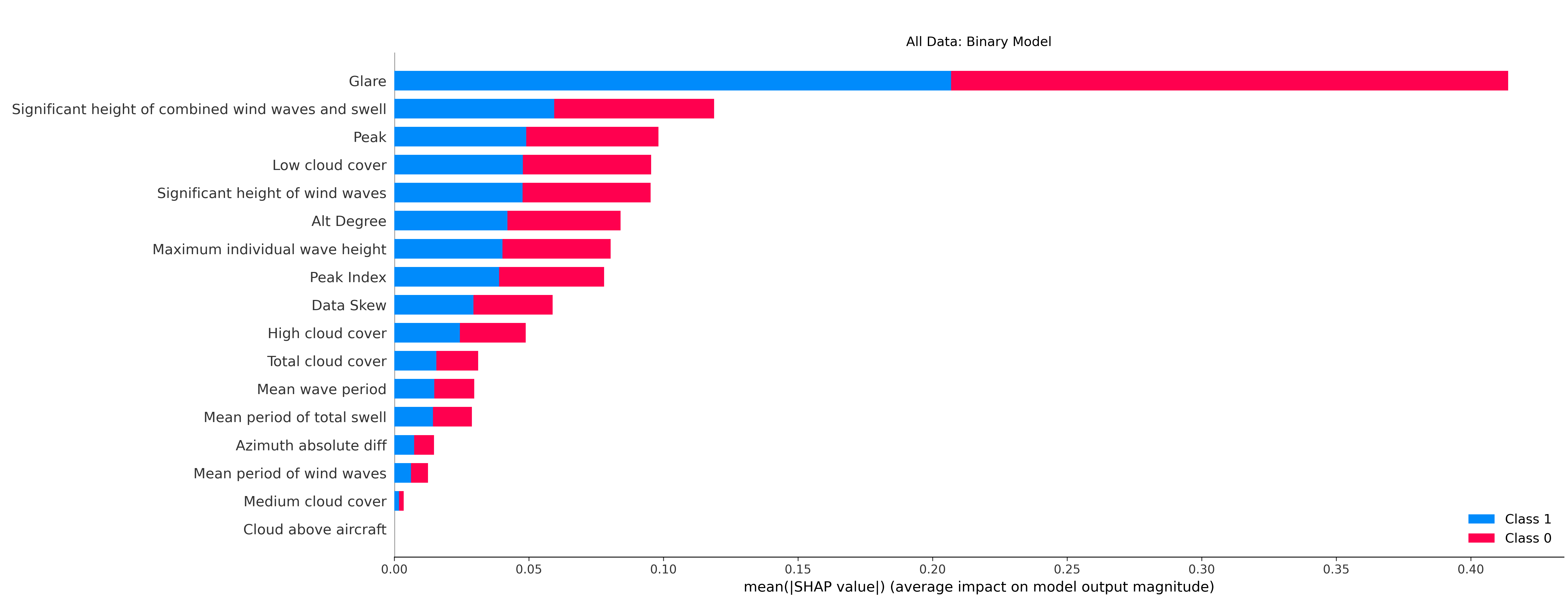}}\\
\vspace{0.1cm}
\fbox{\includegraphics[width=.95\linewidth]{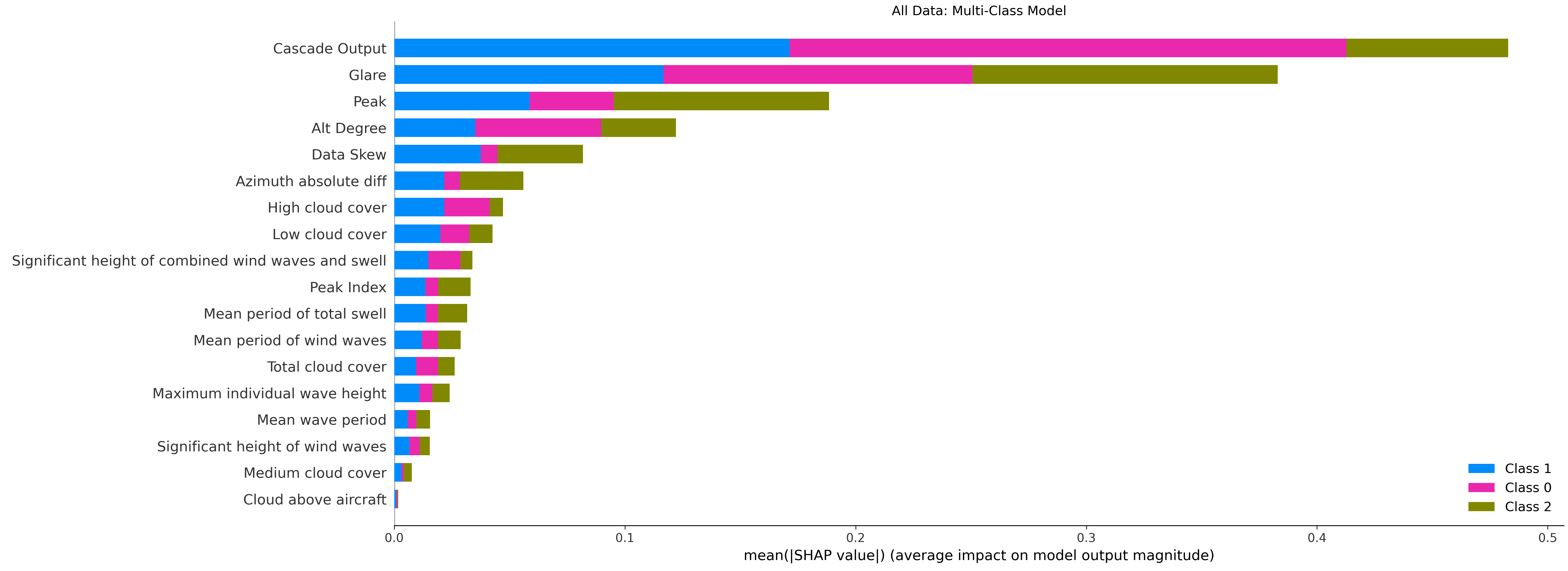}}
\end{center}

   \caption{
   \label{fig:CRFM_shap_cvaui} Summary plot of SHapely values CRFM \textbf{(No GPS Alignment)}. \textbf{Top}: Binary Classification Stage, where Class: 0 is \textit{None} and Class: 1 is \textit{Intermediate} and \textit{Severe}. \textbf{Bottom:} Final Classification Stage, where Class: 0 is \textit{None}, Class: 1 is \textit{Intermediate}, and Class: 2 is \textit{Severe}. 
   }
\end{figure*}

\begin{figure*}[ht]
\begin{center}
\fbox{\includegraphics[width=.95\linewidth]{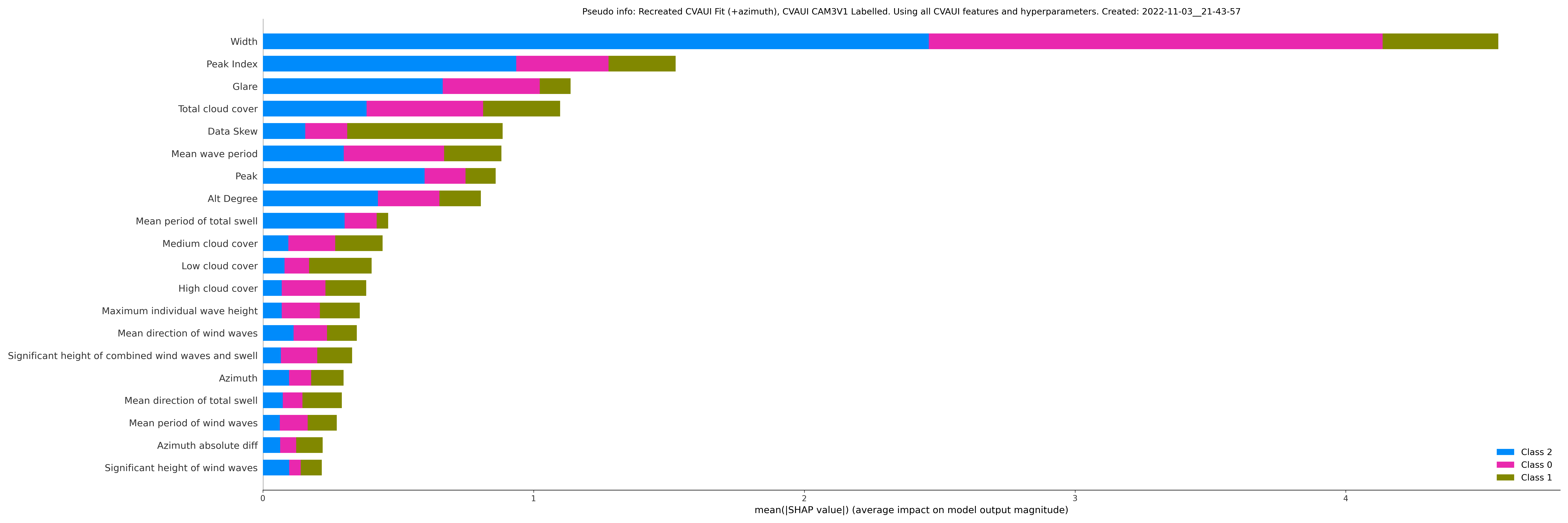}}\\
\vspace{0.1cm}
\fbox{\includegraphics[width=.95\linewidth]{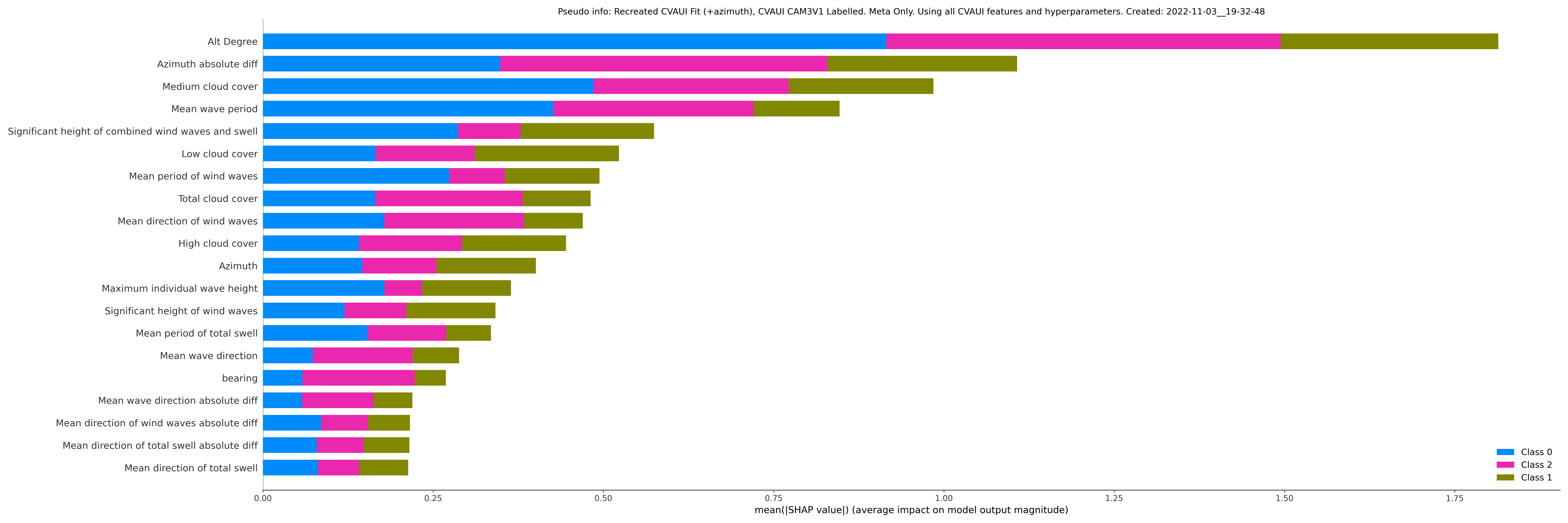}}
\end{center}

   \caption{
   \label{fig:MLP_shap_cvaui} Summary plot of SHapely values for MLP \textbf{(No GPS Alignment)}. \textbf{Top:} Classification MLP, $g_{tc}$, where Class: 0 is \textit{None}, Class: 1 is \textit{Intermediate}, and Class: 2 is \textit{Severe}. \textbf{Bottom:} Prediction MLP, $g_{tp}$, where Class: 0 is \textit{None}, Class: 1 is \textit{Intermediate}, and Class: 2 is \textit{Severe}.
   }
\end{figure*}

SHapley Additive exPlanations (SHAP)\cite{shap2017} is an explainable AI tool which uses game theory to explain the output of machine learning models. It can be used to observe how each feature fed into a model contributes toward classification. The summary plot shown in Figure~\ref{fig:CRFM_shap} illustrates how each stage of our CRFM is interacting with a number of features. Features with bigger contributions appear at the top of the summary plot and have larger SHAP values whereas features that contribute little towards classification have smaller SHAP values and appear at the bottom of the plot (smallest contributors are dropped from the plot altogether). Figure~\ref{fig:MLP_shap} illustrates how the same features fed into CRFM effect classification and prediction in MLP. In our previous paper \cite{cvaui_2022_placeholder} the orientation of the aircraft is not used in the CRFM to attempt to mitigate inevitable orientation errors caused by the small size of the FIT dataset. Figure~\ref{fig:CRFM_shap_cvaui} illustrates how features would have contributed to our previous CRFM had we used aircraft orientation without the proposed GPS alignment methodology shown in this paper. Likewise, Figure~\ref{fig:MLP_shap_cvaui} depicts how our previous MLP utilizes features and again undergoes no GPS-alignment process.

\section{Discussion and Conclusion}
\label{sec:conc}

This manuscript presents a new GPS alignment approach which makes use of two different GPS signals to account for sampling and measurement errors. Embedding this methodology into our existing pipeline \cite{cvaui_2022_placeholder} results in more stringent acceptance levels for data used in our glare classification/prediction systems. A clear distinction between model performance before and after aligning GPS data is visible in both classification performance and prediction performance. Table~\ref{tab:res_class} which illustrates classification performance, shows a significant improvement in accuracy from 92.94\% to 96.22\% when flight tracks are aligned. It also sees a significant improvement in None Class precision from 95.82\% to 99.02\%, which has been identified as a particularly important performance metric \cite{powerClassifyingGlareIntensity2021}, as misclassification of \textit{None} class has the highest consequence on the detection function. Although no special mechanisms are used to increase \textit{None} class precision, the model consistently performs well with this performance metric. This is likely due to the presence of pseudo-labelling bias from CRFM which is designed to perform specifically well on \textit{None} precision. Likewise, in Table~\ref{tab:res_pred}, prediction performance increases with the addition of GPS-aligned data. Prediction sees a substantial improvement from 89.22\% accuracy to 94.01\% accuracy and a \textit{None} class precision improvement from 94.89\% to 98.90\%. 
 
The explainable AI tool SHAP was used to crack open models in our pseudo-labelling stage as well as our semi-supervised learning stages to observe if there are biases that might exist between stages. Figure~\ref{fig:CRFM_shap} illustrates how features interact with the CRFM model in both stages. The binary classification stage's most dominant features are \textit{Glare}, \textit{Alt Degree}, and \textit{Azimuth absolute diff}. Whereas the final classification stage sees \textit{Cascade Output}, \textit{Glare}, and \textit{Peak} as its most dominant features. Moving on to classification and prediction MLP models in Figure~\ref{fig:MLP_shap} we see a different order of feature importance. \textit{Width}, \textit{Alt Degree}, and \textit{Azimuth absolute diff} are identified as classifications most contributory features. On the other hand, prediction's largest contributors are \textit{Alt Degree}, \textit{Azimuth absolute diff}, and \textit{Medium cloud cover}. The difference in feature importance between CRFM and MLP models indicates that MLP models are not trying to mimic the CRFM architecture used to create pseudo-labels, but rather they are trying to learn how to best classify/predict glare intensities. It is also apparent that when presented with imagery-related features models will choose at least one of $\cal F$ features as a dominant contributor. Outside of imagery-related features \textit{Alt Degree} and \textit{Azimuth absolute diff} are continuously ranked as large contributing features. Additionally, SHAP summary plots are also shown for our previous models, exposed to data without GPS alignment. Figure~\ref{fig:CRFM_shap_cvaui} and Figure~\ref{fig:MLP_shap_cvaui} illustrate how CRFM and MLP, respectively, utilize features toward classification/prediction. Comparing CRFM feature importance from this manuscript, Figure~\ref{fig:CRFM_shap}, and that of our previous paper, Figure~\ref{fig:CRFM_shap_cvaui}, we see a significant change in the utilization of features in the binary stages of each CRFM but little change in the final stages. The binary stage of the non-GPS-aligned model ranks the importance of \textit{Alt Degree} and \textit{Azimuth abs diff} far lower than its GPS-aligned counterpart. This is likely because, as anticipated, the sparseness of the FIT dataset is not sufficient to extract orientation without help from flight tracks (O-GPS, signals). Comparing MLP feature importance between this manuscript, Figure~\ref{fig:MLP_shap}, and our previous paper, Figure~\ref{fig:MLP_shap_cvaui}, it is observed that features are used differently between classification and prediction in both models. When observing classification, the top portions of each figure, the GPS-aligned version relies on imagery and orientation most heavily for classification. However, the non-GPS-aligned version relies primarily on imagery. When observing prediction, the lower portions of each figure, it is shown that feature importance is quite similar between the two models once imagery features are removed. However, the SHAP value for the orientation feature \textit{Azimuth abs diff} is more than two times smaller in the non-GPS-aligned model when compared to the GPS-aligned model. In summary, utilizing our proposed GPS alignment methodology changed the network behaviour of our architecture. This change in behaviour increased model confidence and reliance on features identified as imperative in our image formation model. 

By design, our naïve pseudo-labelling approach filters out data known to present uncertainty through the use of confidence tolerances \cite{cvaui_2022_placeholder}. In addition to this, our proposed GPS alignment algorithm is also used to mitigate uncertainty by relying on weighted centroids which are synthetically interpolated points $SI$, which, in effect remove geotag outliers. The rejection analysis undertaken in this manuscript was aimed at investigating uncertainty in our pre-processing and dataset generation stages and observing if rejected labels consisted of instances of poor data quality. Of the many spatiotemporal proximity analyses, only one yielded meaningful results, shown in Figure~\ref{fig:reject}. It is observed that the median value for the number of nearby I-GPS points between all three clusters (All data, accepted pseudo-labels, and rejected pseudo-labels) remains the same. However, when observing the distributions of the more accurate O-GPS points a median count of 7 is present in the all data and accepted pseudo-label clusters whilst a value of 4 is observed in the rejected pseudo-label cluster. Therefore, a given synthetically interpolated point, $SI_i$ is deemed more reliable when the points surrounding it, $P_n$, have a higher concentration of O-GPS points. Meaning, the increased value for $W_{source}$ is justified and can be used to remove I-GPS outliers which sometimes hamper accurate orientation measures. Additionally, these findings bode well with how one might expect uncertainty to present itself in our models, which can be boiled down to two main culprits: the presence of land in imagery and inaccurate global positioning (since all features outside of $\cal F$ rely on the coordinates of the aircraft). In an attempt to mitigate land in a scene which introduces uncertainties in $\cal F$, a high-resolution land-sea mask is utilized. However, this mask relies on accurate aircraft position. Therefore, in effect uncertainty is mostly due to inaccurate positioning which relies on I-GPS and O-GPS geotags, O-GPS being the more accurate of the two. Given most navigational devices come pre-equipped with GPS sensors our proposed alignment methodology could be applied to a variety of applications, specifically, applications where orientation is of key importance. This is the case in nearly all optical imaging systems when hoping to achieve optimal data collection. Using this approach poses a cost-effective means of determining orientation without the need for additional sensors. 

The objective of this research is to produce a data-driven mission planning system which uses meteorological and astronomical data to inform surveyors of optimal flight trajectories in order to cover the most area as effectively as possible. Using such a system mitigates errors that are often viewed as occupational hazards and allow for the acquisition of larger quantities of useful data. Search and rescue is one of a large variety of useful applications. Here, planning search transects to improve subsurface visibility could dramatically increase one's ability to detect missing persons promptly. Likewise, this could be applied to subsurface applications where objects are near the surface of the water but might be difficult to detect with poor acquisition angles. An example of this is submarine detection when vessels are approaching a breach or beginning to dive.

 % In navigational applications where magnetometers are used, it is known that Earth's magnetic field is not uniform, and this can affect heading readings \cite{jiaUseGPSSensors2004}. In areas known to be affected by nefarious magnetic fields, GPS alignment could be used in its stead.

\section{Future Work}
\label{sec:future}
Preemptive path planning to optimize survey visibility through data-driven mission planning depends on the generalizability of existing models. For future work, we propose holding out flights as a validation set to observe how well our prediction model performs along a historical flight track. Once a sufficient performance is achieved, a path planning algorithm should be developed using said prediction model on yet another historical flight to observe if the presence of glare was reduced. Once this milestone is achieved data-driven mission planning capabilities will be at a level ready for testing in real-world settings. This research continues to strive toward developing a comprehensive understanding of circumstances which hinder subsurface visibility and survey effectiveness. Understanding and mitigating these effects remain at the forefront of this project's guiding principles as it justifies a means of providing accurate population estimates that contribute to the conservation and longevity of critically endangered species.

\section*{Acknowledgments}
The authors would like to thank both Stephanie Ratelle and Elizabeth Thompson for granting access to the CAM3 dataset and its associated flight tracks. The authors would also like to thank Mylene Dufour, Marie-France Robichaud, and Maddison Proudfoot for their dedicated work in preparing and labelling the FIT dataset. Their experience in conducting methodical aerial surveys throughout Atlantic Canadian waters was essential to our work.

\bibliographystyle{IEEEtran}
\bibliography{bibliography}
\end{document}